\documentclass[journal=nano,manuscript=article]{achemso}

\usepackage[T1]{fontenc} 
\usepackage{amsmath}    
\usepackage{graphicx}   
\usepackage{verbatim}   
\usepackage{color}      
\usepackage{subfigure}  
\usepackage{hyperref}   
\usepackage{gensymb}
\usepackage{xcolor}

\title{\textcolor{black}{Monolithic} 3-Dimensional Tuning of an Atomically Defined Silicon Tunnel Junction}
\author{Matthew B. Donnelly}
\affiliation{Centre for Quantum Computation and Communication Technology, School of Physics, University of New South Wales, Sydney 2052, NSW Australia}
\author{Joris G. Keizer}
\affiliation{Centre for Quantum Computation and Communication Technology, School of Physics, University of New South Wales, Sydney 2052, NSW Australia}
\author{Yousun Chung}
\affiliation{Centre for Quantum Computation and Communication Technology, School of Physics, University of New South Wales, Sydney 2052, NSW Australia}
\author{Michelle Y. Simmons}
\affiliation{Centre for Quantum Computation and Communication Technology, School of Physics, University of New South Wales, Sydney 2052, NSW Australia}
\email{m.donnelly@unsw.edu.au}

\begin{document}

\begin{abstract}
A requirement for quantum information processors is the in-situ tunability of the tunnel rates and the exchange interaction energy within the device. The large energy level separation for atom qubits in silicon is well suited for qubit operation but limits device tunability using in-plane gate architectures, requiring vertically separated top-gates to control tunnelling within the device. In this paper we address control of the simplest tunnelling device in Si:P – the tunnel junction. Here we demonstrate that we can tune its conductance by using a vertically separated top-gate aligned with +-5nm precision to the junction. We show that a \textcolor{black}{monolithic} 3D epitaxial top-gate increases the capacitive coupling by a factor of 3 compared to in-plane gates, resulting in a tunnel barrier height tunability of 0-186meV. By combining multiple gated junctions in series we extend our \textcolor{black}{monolithic} 3D gating technology to implement nanoscale logic circuits including AND and OR gates.
\end{abstract}

\newpage

Electron spins are a leading candidate for qubits in solid-state quantum computing. To form such qubits it is necessary to spatially confine a small number of electrons to form a quantum dot, which can be achieved either at a hetero-interface using metallic surface gates \cite{Kouwenhoven1991, Kawakami2014, Veldhorst2014} or using the Coulomb potential of an \textit{n}-type donor \cite{Pla2012, Fuechsle2012}. \textcolor{black}{In the original proposal for a donor-based quantum computer by Kane \cite{Kane1998} additional metallic surface gates were envisioned above a dielectric to control the inter-donor tunnelling between phosphorus qubits. However, the large thermal budget (T $>$ 700$\degree$C \cite{chabal2001}) required to produce a dielectric with low defect density causes diffusion of the STM-patterned phosphorus-doped structures to the extent that the device is inoperable.} \textcolor{black}{Instead we find that nanoscale gates patterned by STM lithography \cite{OBrien2001, Ruess2004, Shen1995, Lyding1994} in an all epitaxial approach provide excellent control circuitry for the qubits and can reach the strong response regime required for fast, high fidelity operation \cite{Keith2019}.} This atomically precise doping technique has the added advantage that the active region of the device is separated away from any surfaces or interfaces, resulting in a low noise environment \cite{Kranz2020, Shamim2016}. To date critical components for a scalable quantum computer have been realised using this monolithic doping method such as metallic nanowires \cite{Weber2012}, single atom transistors \cite{Fuechsle2012}, single electron transistors (SETs) \cite{Keith2019}, single spin control \cite{Hile2018} and a two-qubit gate \cite{He2019}. The atomic precision with which these components can be made, and their small size, can be leveraged for scalable quantum computing architectures such as linear arrays of singlet-triplet qubits \cite{Pakkiam2018} or for single spin qubits in a 2D surface code design \cite{Hill2015}. Ultimately, error correction requires realisation of the 2D surface code with multi-layer STM lithography with control electrodes above and below the 2D qubit array \cite{McKibbin2013, Koch2019}. The ability to place nanoscale control gates with atomic precision in close proximity to the qubits but vertically separated in different crystalline planes allows us to tune the exchange coupling \textit{\textbf{J}} and hyperfine value \textit{\textbf{A}} as originally conceived by Kane.

Central to all nanoscale devices are the junctions between device elements where electrons have to tunnel from one electrode to another. In atomically precise qubit devices these tunnel junctions exist either between the qubits themselves - typically on the order of $\sim$10-15\,nm or between the qubit and the readout sensor ($\sim$20\,nm). Tunnel junctions can also be formed between the terminated ends of two nanowires, and form a key component in qubit readout applications such as in single electron transistors (SETs) \cite{Keith2019}, in single-lead quantum dots \cite{House2016} and in charge sensing tunnel junctions \cite{House2014}. In plane gates are capacitively coupled to different elements within the device, and depending on the distance to the gates (typically 50-100\,nm) the lever arm $\alpha$ (\textit{i.e.} the ratio of the potential change at the active region of the device to the potential change on the gate) is typically $\sim$0.1. The ability to create vertically separated gates for single shot spin readout has shown a larger $\alpha\sim0.3$ for gates separated by 100\,nm \cite{Koch2019} with this increased coupling motivating further work to investigate 3D epitaxial top gates, particularly with respect to controlling the tunnelling of electrons in Si:P nanoscale devices. \textcolor{black}{In this work we leverage the increased coupling of 3D epitaxial gates to control tunnelling in Si:P tunnel junctions, providing a functionally useful component in fabricating Si:P quantum processors. We demonstrate that this technique can be used for on-chip logic and has the potential to facilitate qubit control and readout schemes.}

\begin{figure}[]
	\includegraphics[width = 8.5cm]{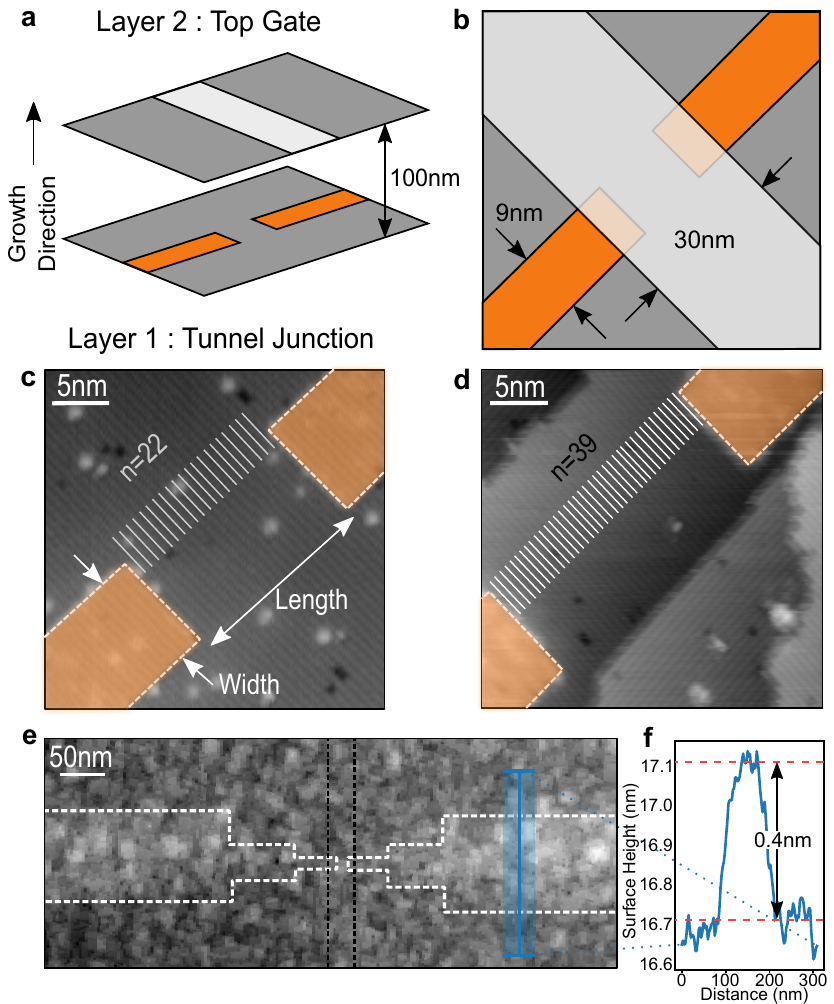}
	\caption{\textbf{Precision alignment of a 30\,nm wide epitaxial top gate above a 8.8\,nm$\times$16.9\,nm and 10.0\,nm$\times$30.0\,nm Si:P tunnel junction.} \textbf{a.} An illustration of a vertically gated 3D tunnel junction in Layer 1 (orange leads) separated from the top-gate in Layer 2 (grey lead) by 100\,nm of epitaxially grown silicon. \textbf{b.} A top-down illustration of the device showing the $\sim$5\,nm alignment accuracy of the top-gate in Layer 2 with respect to the tunnel junction in Layer 1. \textbf{c.} An STM image of the tunnel junction with a short (\,$n$\,=\,22 dimer rows = $16.9$\,nm $\pm$ 0.8\,nm) channel length, with the bright outlined areas indicating the desorbed regions where phosphorus atoms will be incorporated into the silicon lattice. \textbf{d.} An STM image of a long channel length junction separated by \,$n$\,=\,39 dimer rows, giving a junction length $ L=n\times0.768$\,nm\,$=30.0$\,nm\,. \textbf{e.} An STM image of the surface of Layer 2 showing the topographic outline of the device on Layer 1 (white dashed line) surviving encapsulation with 100\,nm of silicon. The 30\,nm wide top-gate (black dashed line) is aligned directly over the tunnel junction with alignment accuracy of $\pm$5\,nm. \textbf{f.} An averaged line profile from the STM image in \textbf{e} of Layer 2, showing the $\sim$0.4\,nm difference in surface height caused by the phosphorus doped region of Layer 1.}
	\label{fig1}
\end{figure}

We first demonstrate control of the barrier height in single planar tunnel junctions using a vertically separated top-gate within a monolithic device, illustrated in Figure \ref{fig1}a, b. On the first lithographic plane (Layer 1) a tunnel junction is fabricated with a width $W$ and length $L$, and in Figure \ref{fig1}c, d we show images of two different tunnel junctions fabricated for this work with dimensions 8.8\,nm$\times$16.9\,nm and 10.0\,nm$\times$30.0\,nm, \textcolor{black}{with all dimensions having an uncertainty of $\pm$0.8\,nm due to the chemistry of donor incorporation \cite{Fuechsle2012, Wilson2006}}. Fabrication of the 3D-gated tunnel junctions is performed using scanning tunnelling microscopy (STM) hydrogen resist lithography \cite{OBrien2001, Simmons2003, Shen1995, Lyding1994}, where an STM tip is used to selectively desorb the hydrogen mask bound to a 2$\times$1 reconstruction of the (001) silicon surface. This process is performed in an ultra-high vacuum (UHV) system with a base pressure of $\sim$1$\times$10$^{-11}$\,mbar. After lithography we dose the surface with phosphine gas (PH$_3$) at a pressure of $5\times10^{-7}$mbar for 2 minutes (1.8 Langmuir) and incorporate the phosphorus atoms into the exposed silicon using a 350\degree C anneal for 60 seconds. Subsequent to doping on Layer 1, a 100\,nm layer of silicon is grown at a temperature of 250\degree C and a growth rate of 0.15\,nm/min using kinetic growth manipulation (KGM) \cite{Esser2004, Koch2019}. Here a sequence of four rapid thermal anneals to 450\degree C for one minute each, spaced 5 minutes apart, are executed in the last phase of growth. This process minimises the roughness of the resulting surface making it suitable for subsequent STM lithography. On the second lithographic plane (Layer 2) a 30\,nm wide top gate is fabricated directly over the tunnel junction. The positioning of the top gate is possible by aligning to the surface height topography of the doped regions of Layer 1. Here the doped regions appear higher ($\sim$0.4\,nm) topographically and can be observed when imaging the surface of Layer 2 (see the marked dashed white outline in Figure \ref{fig1}e and averaged line cut in Figure \ref{fig1}f). \textcolor{black}{The reason for this height difference is that the 350\degree C incorporation anneal causes phosphorus to substitute for silicon into the top layer of the surface leading to ejected silicon and fragments of PH$_x$ and H remaining on the surface before silicon growth. These species impact the growth of silicon at 250ºC and therefore modify the topography of the overgrown surface \cite{McKibbin2009}. We believe it is the presence of a low density of hydrogen atoms and fragments of PH$_x$ in the dosed region which contribute to changes in the growth quality and therefore to the height difference observed on the overgrown layer.} To electrically contact the device, silicon vias aligned with STM-defined contact pads are etched using reactive ion etching and aluminium is used to make Ohmic contact with the phosphorus-doped silicon layer. 

\begin{figure}[]
	\includegraphics[width = 8.5cm]{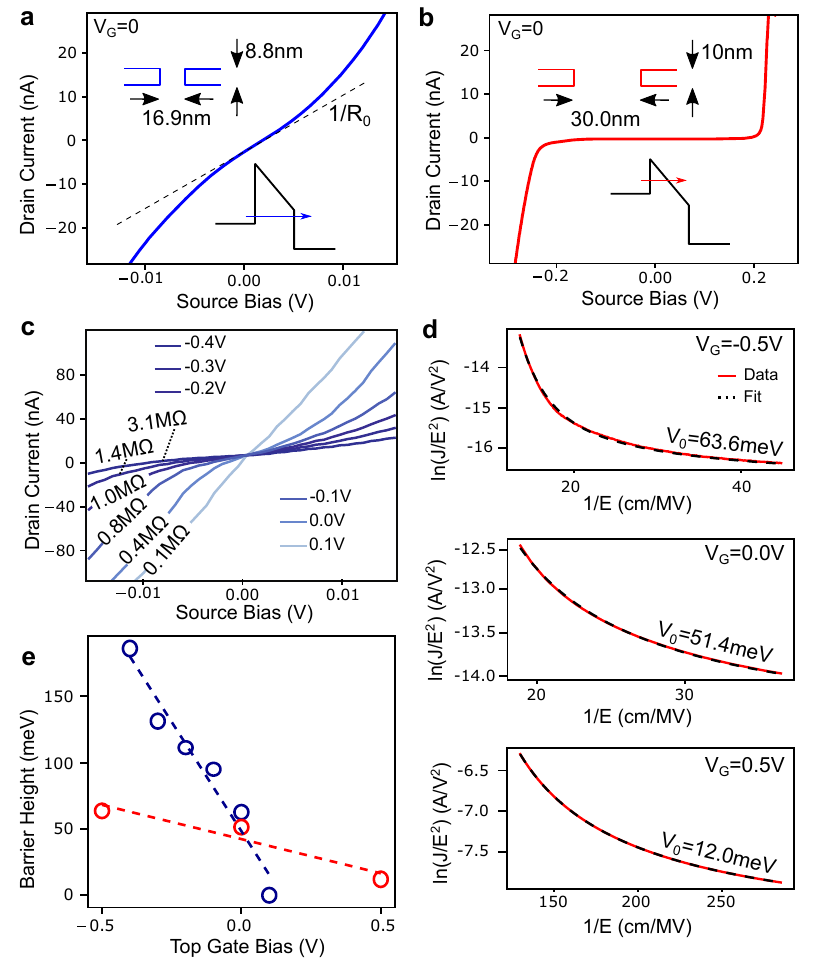}
	\caption{\textbf{Control and measurement of the tunnel junction barrier height.} \textbf{a.} An \textit{IV}-\,curve measured on the short-barrier junction, with the zero-bias gradient indicated by the dashed line. The lower inset shows a sketch of the energy levels in the junction, indicating tunnelling through the full length of the trapezoidal barrier. The upper inset indicates the dimensions of the junction. \textbf{b.} \textit{IV}-\,curve of a long-barrier (30.0\,nm) junction showing suppression of the current at low bias. The lower inset shows a sketch of the energy levels in the junction indicating tunnelling through a triangular barrier, with the upper inset indicating the dimensions of the junction. \textbf{c.} Source-drain curves as a function of top-gate bias $V_G$ for the short-barrier (16.9\,nm) junction with $R_0$ labelled for each curve. \textbf{d.} Extraction of tunnel junction barrier heights with fits (black dashed line) to the Fowler-Nordheim equation for a long barrier junction. \textbf{e.} Barrier heights as a function of $V_g$ extracted from the source-drain curves for the short-barrier junction (direct tunnelling WKB, blue) and long-barrier junction (Fowler-Nordheim, red).}
	\label{fig2}
\end{figure}

To characterise the conductance through our two tunnel junction devices we measure them in a two-terminal configuration at 4.2K. Here the intrinsic silicon substrate becomes insulating due to charge freeze-out and the degenerately phosphorus-doped regions are quasi-metallic due to multiple bands being pulled below the Fermi energy by the phosphorus doping \cite{Ryu2013}. Conduction through the tunnel junction is primarily determined by two parameters: the length of the junction $L$ and the height of the potential barrier $V_0$. Depending on $L$, conduction through the junction will either be dominated by direct tunnelling (DT) or Fowler-Nordheim tunnelling (FNT) \cite{Fowler1928}. Direct tunnelling occurs through a trapezoidal barrier where a source-drain \textit{IV}-curve is characterised by a finite zero-bias gradient at low bias, see Figure \ref{fig2}a. In contrast Fowler-Nordheim tunnelling occurs at higher biases through a triangular barrier that is modified by the electric field $E$ across the barrier \cite{Depas1995} (Figure \ref{fig2}b). For this work the tunnel junction with a shorter barrier (Figure \ref{fig1}c) is in the direct tunnelling regime, and the tunnel junction with a longer barrier (Figure \ref{fig1}d) shows no direct tunnelling at low bias and is in the Fowler-Nordheim tunnelling regime at high bias ($V_{S}>$0.2\,V).

To tune the barrier height $V_0$ in each device we apply a gate voltage $V_G$ to the top-gate patterned on Layer 2. \textcolor{black}{We assume the top gate does not change the electrostatic dimensions of the tunnel junction as the large number of phosphorus nuclei present from the high doping density \cite{Shen2002, Oberbeck2002, Goh2006} ($\sim$\,80 in the last 5\,nm of the leads) will provide a deep Coulomb-like potential that strongly confines the electronic wavefunction of the junction. Such an assumption has been used previously in modelling of Si:P nanostructures \cite{Ryu2015}, and is used in studies of tunnel junctions in MOSFETs \cite{Shirkhorshidian2018} despite the  weaker confinement provided in the inversion channel of a MOSEFT compared to the strong Coulomb potentials of the phosphorus donors in our Si:P tunnel junctions.} Consequently on application of a voltage to the top gate we expect only the barrier height $V_0$ to change with gate bias $V_G$. In the case of the short barrier length we extract a barrier height $V_0$ from the zero-bias resistance $R_0$ measured from the $IV$ data (See Figure \ref{fig2}c) using the WKB approximation for direct tunnelling resistance \cite{Wang2020}
\begin{equation}
R_0 = \frac{h}{2e^2}e^{\frac{2L}{h}\sqrt{2mV_0}}. 
\end{equation}
In the short barrier length junction with the highest top-gate bias ($V_G$=0.1\,V) we see that there is no curvature of the source-drain trace, which indicates that there is no tunnelling transport taking place and that conduction is Ohmic. Furthermore, the resistance of the curve (0.1M$\Omega$) is within the range of contact resistances (50k$\Omega$-100k$\Omega$) and lead resistances (20k$\Omega$-40k$\Omega$) we typically expect in Si:P devices. We therefore conclude that for $V_G$=0.1V the barrier in the shorter tunnel junction is completely lowered, and that the resistance of 0.1M$\Omega$ we measure is the combined contact resistance and lead resistance. To accurately calculate the barrier heights using the WKB equation we normalise our measured resistances by subtracting the combined contact and lead resistances of $\sim$0.1M$\Omega$ from our measured resistances.

For the long channel length junction we extract $V_0$ by fitting the $IV$ data to an analytic expression for the FNT current density, $J(E)$: 
\begin{equation}
J(E) = AE^2e^{(-\frac{B}{E})}, 
\end{equation}
where $A=\frac{me^3}{8\pi h m^* V_0}$, $B=\frac{8\pi\sqrt{2m^*}V_0^{3/2}}{3he}$ and $E$ is the electric field across the junction \cite{Miranda2011}, see Figure \ref{fig2}d. Barrier heights as a function of $V_G$ for both junctions are plotted in Figure \ref{fig2}e. We note that at $V_G$ = 0\,V we extract barrier heights of 62.6\,meV$\pm$5\,meV and 51.3\,meV$\pm$10\,meV for the short and long barrier junctions respectively, in agreement with recently reported experimental values ($V_0$=100\,meV$\pm$50\,meV \cite{Wang2020}) for $V_0$ in Si:P tunnel junctions in the direct tunnelling regime ($W\sim$11.5\,nm and L$<$12.3\,nm). Only the zero gate bias $V_G=0$ values of $V_0$ in our work can be compared to the values extracted in Wang \textit{et. al.} \cite{Wang2020} as no gating, either 2D or 3D, was reported in that study.  
 
On biasing the top gate we demonstrate that we can tune the barrier height $V_0$ by 186\,meV and 52.6\,meV for the short and long barrier devices respectively. This equates to lever arms $\alpha$ of 0.37 and 0.05 respectively. For reference, lever arms of 0.1 are typical for in-plane Si:P gates and a lever arm of 0.3 has been demonstrated with a 3D epitaxial gate by Koch \textit{et. al.} \cite{Koch2019}. \textcolor{black}{Using the WKB approximation, we find that a change in the barrier height of 186\,meV will modulate the tunnel coupling between donor atom qubits by a factor of $\sim$30, providing us with another means to control the exchange coupling in donor atom two-qubit gates which currently solely rely on in-plane gate \cite{He2019}}. The smaller tunability of the long channel device is likely due to the top gate being of a similar width to the channel and hence not gating the barrier uniformly, while the lever arm in the short channel device is 25\% larger than previously demonstrated 3D epitaxial gates \cite{Koch2019}. The tunability of future 3D-gated tunnel junctions can be further increased by bringing the top-gate closer to the tunnel junctions by reducing the height of the silicon layer between Layer 1 and Layer 2 \textcolor{black}{while maintaining minimal leakage current between the 3D epitaxial gate and the layer below. For reference, the leakage current in the devices measured in this work is $<$100\,pA across the full gate range shown.}

\begin{figure}[t]
	\includegraphics[width=11.8cm]{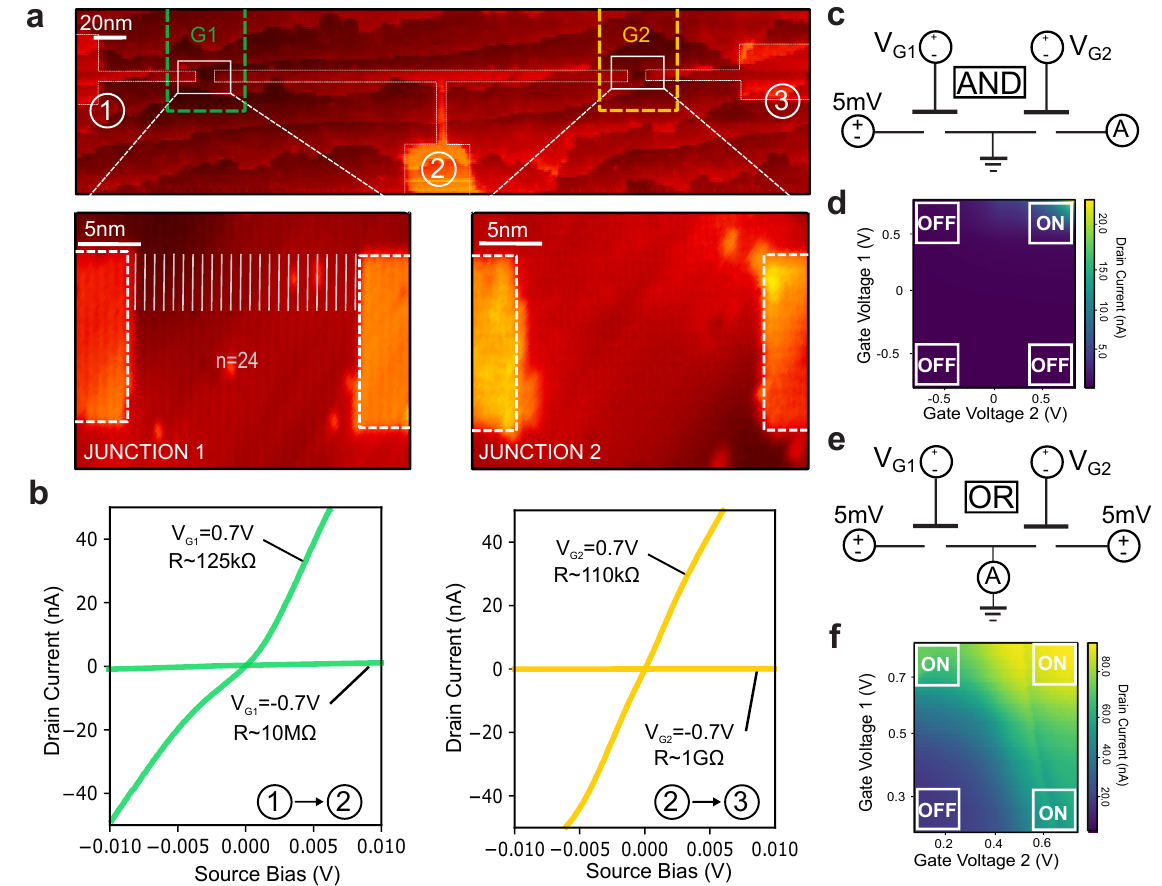}
	\caption{\textbf{Logic operations in a three terminal tunnel junction device.} \protect \textbf{a.} (top) \textcolor{black}{Constant-current} STM images of the inner part of the device indicating three terminals (\raisebox{.5pt}{\textcircled{\raisebox{-.9pt} {1}}}, \raisebox{.5pt}{\textcircled{\raisebox{-.9pt} {2}}}, \raisebox{.5pt}{\textcircled{\raisebox{-.9pt} {3}}}) used to measure two gated tunnel junctions (white boxes), and the alignment of the top-gates G1 (green-dashed) and G2 (yellow-dashed). \textcolor{black}{The light-coloured regions indicate where the hydrogen termination has been desorbed by the STM tip, revealing regions of bare silicon dangling bonds.} \textbf{b.} Source-drain $IV$-curves of Junction 1 (\raisebox{.5pt}{\textcircled{\raisebox{-.9pt} {1}}}$\,\to\,$\raisebox{.5pt}{\textcircled{\raisebox{-.9pt} {2}}}, green) and Junction 2 (\raisebox{.5pt}{\textcircled{\raisebox{-.9pt} {2}}}$\,\to\,$\raisebox{.5pt}{\textcircled{\raisebox{-.9pt} {3}}}, yellow) as their respective top-gates G1 and G2 are swept from -0.7V to 0.7V. \textbf{c. e.} AND and OR logic circuits. \textbf{d. f.} Maps of the measured drain current (nA) as a function of gate voltages $V_{G1}$ and $V_{G2}$. The white boxes in the corners of the maps represent the regions of high/low current which correspond to the ON/OFF logic outputs for each operation.}
	\label{fig3}
\end{figure}

We now consider the potential of 3D top-gated tunnel junctions for the implementation of transistor-like operation at cryogenic temperatures. The concept of harnessing quantum tunnelling as a switching mechanism in a gated field-effect transistor (FET) has been proposed and implemented in various material platforms such as Schottky barrier CMOS \cite{Tucker1994}, carbon nanotubes \cite{Heinze2002}, and van der Waals heterostructures \cite{Xiong2020}, with the primary benefit being that tunnelling conduction is not limited by the thermal Maxwell-Boltzmann tail, leading to higher current on/off ratios and lower power consumption \textcolor{black}{at room temperature. Si:P tunnel junctions provide a unique platform for investigating this concept as their atomic precision, sharp dopant profiles, and monolithic structure means that they can be integrated at the device level within a larger epitaxial Si:P structure.}

To demonstrate transistor-like operation based on tunnelling we pattern two gated tunnel junction in series with a second terminal connected between them, as shown in Figure \ref{fig3}a, with both junctions patterned with dimensions $W$ = 8.5\,nm  and $L$ = 18.4\,nm, placing them in the direct tunnelling regime. This was deliberately chosen since longer junctions would require operation at relatively high source-drain biases ($>$200\,mV) which is undesirable due to heat dissipation and leakage. We can characterise each tunnel junction individually by taking $IV$-curves as a function of their respective gate voltages $V_{G1}$ and $V_{G2}$ between -\,0.7\,V and 0.7\,V, using contacts \raisebox{.5pt}{\textcircled{\raisebox{-.9pt} {1}}}$\,\to\,$\raisebox{.5pt}{\textcircled{\raisebox{-.9pt} {2}}} and \raisebox{.5pt}{\textcircled{\raisebox{-.9pt} {2}}}$\,\to\,$\raisebox{.5pt}{\textcircled{\raisebox{-.9pt} {3}}}. Figure \ref{fig3}b shows this data, demonstrating that both tunnel barriers can be raised and lowered with their respective top gates G1 and G2, and with current on/off ratio $I_{on/off}$=1000 at a source bias of $V_S$=5\,mV. \textcolor{black}{The difference in zero-bias resistance of both junctions, with nominally the same dimensions, but with $\pm$0.8\,nm uncertainty due to the exact location of the dopant after the incorporation anneal, is most likely due to variations in the density of states of the leads that are known to be sensitive to small variations in the lead width. \cite{Ryu2013, Ryu2015}} 

We can now operate this device in two modes, as an AND gate or an OR gate, to demonstrate two separate logic operations (see Figure \ref{fig3}c, e). First we consider an AND gate by applying 5\,mV to terminal \raisebox{.5pt}{\textcircled{\raisebox{-.9pt} {1}}} and measuring the current at the terminal \raisebox{.5pt}{\textcircled{\raisebox{-.9pt} {3}}}, see Figure \ref{fig3}c. A map of the current as a function of the gate voltages $V_{G1}$ and $V_{G2}$ is shown in Figure \ref{fig3}d with the white boxes in the corners of the map indicating the logical states represented by the high/low states of the current. The high current ON state was observed at a source bias $V_S=5$\,mV to be approximately 25\,nA in agreement with the $IV$ data in Figure \ref{fig3}b. At $V_S$=5\,mV the ON state resistance of each junction was approximately 0.1M$\Omega$ giving a total series resistance of 0.2M$\Omega$ resulting in our measured current of 25\,nA. The low current state was $<$50\,pA at $V_S$=5\,mV, resulting in a current on/off ratio $I_{on/off}=500$. 

Next, we construct an OR gate by applying 5\,mV to both terminals \raisebox{.5pt}{\textcircled{\raisebox{-.9pt} {1}}} and \raisebox{.5pt}{\textcircled{\raisebox{-.9pt} {3}}} and measuring the current at terminal \raisebox{.5pt}{\textcircled{\raisebox{-.9pt} {2}}} while sweeping $V_{G1}$ and $V_{G2}$, see Figure \ref{fig3}e and f. The high current state of approximately 100\,nA exceeded that of the AND gate due to bias being applied at both terminals \raisebox{.5pt}{\textcircled{\raisebox{-.9pt} {1}}} and \raisebox{.5pt}{\textcircled{\raisebox{-.9pt} {3}}}. We observe a similar low current state ($<$50\,pA) but get a higher current on/off ratio $I_{on/off}$=2000. 

The on/off ratio of the AND and OR logic gates (500 and 2000 respectively) as well as of the individual tunnel junctions in the logic device (both $\sim$1000) are similar to other experimental transistor platforms such as monolayer graphene ($\sim$1000) \cite{Lu2010}, graphene nanomeshs ($\sim$7000) \cite{Berrada2013}, ferroelectric FETs ($\sim$1000) \cite{Katsouras2015}, and organic FETs ($\sim$8000) \cite{Schroeder2004}, but with the added advantage of being able to operate at temperatures $<$4.2K. Crucially this low temperature operation opens the door for 3D-gated tunnel junctions to provide on-chip logic at the cryogenic temperatures required for the operation of Si:P quantum processors, as peripheral classical logic becomes increasingly important for scaled-up quantum computing architectures \cite{Charbon2016, Incandela2017, Charbon2017}. 

In this work we have fabricated a top-gate on a tunnel junction in a monolithic Si:P architecture and used this 3D gate to demonstrate tunability in barrier height $V_0$ in the range 0-186\,meV, equating to a lever arm of $\sim$0.35 compared to a typical planar-gate lever arm of $\sim$0.1. We demonstrated the operation of a device with two junctions in series as an AND and OR logic gate with current on/off ratios of 500 and 2000 respectively, similar to other nanoscale transistor designs. While this technology may not replace cryo-CMOS technology, it will have applications in quantum computing architectures where small numbers of top-gated Si:P tunnel junctions are integrated within the epitaxial monolithic silicon to provide on-chip logic. The results presented are also a valuable step towards the integration of 3D epitaxial gates in STM devices with potential for control of single electron wavefunctions in donor dot devices to tune tunnel rates, tunnel couplings and hyperfine couplings, facilitating more flexible and precise control in quantum information applications. 

\bibliography{wolf} 

\end{document}